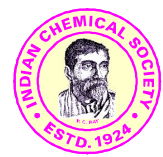

# Utilization of the simulated flue gas on the cultivation of *Chlorella protothecoides*

Cem Özel[a], Muharrem Erdem Boğoçlu[b], Ceren Keçeciler[a], Ecem Kaplan[a,c] and Sevil Yücel[a]*

[a]Faculty of Chemical and Metallurgical Engineering, Department of Bioengineering, Yildiz Technical University, Turkey

[b]Faculty of Mechanical Engineering, Department of Mechanical Engineering,
Yildiz Technical University, Turkey

[c]Faculty of Engineering and Natural Sciences, Department of Bioengineering, Uskudar University,
Istanbul, Turkey

*E-mail:* syucel@yildiz.edu.tr, yuce.sevil@gmail.com

In recent years, fossil-based fuels have been used to supply the energy needs of the world. Fossil-based fuels induce accumulation of the atmospheric $CO_2$ which causes global warming. One of $CO_2$ source is flue gas emission from the power plant. The microalgae have been considered excellent biological materials for reduction $CO_2$ with ability to photosynthesis. In this study, air, 10% $CO_2$, 15% $CO_2$ and simulated flue gas (containing 15% $CO_2$) feed were used to observe effect $CO_2$ concentration and flue gas on cell growth and lipid content of *Chlorella protothecoides*. The highest dry cell weight (1,5 g/L) and lipid content (45%) values were obtained with 15% $CO_2$ feed while the highest growth rate (1,10 μ), biomass productivity (0,125 g/L/day), and lipid weight (0,63 g/g) were observed in 10% $CO_2$ feed. Cultures fed with flue gas did not inhibited *C. protothecoides* growth and showed similar results with those fed with 15% $CO_2$ gas in terms of growth rate, dry cell weight, biomass productivity and lipid content. These results showed that *C. protothecoides* has great potential for reducing $CO_2$ emission from flue gas.

## Introduction

In the past decades, increasing concentrations of carbon dioxide ($CO_2$) and other greenhouse gases in the atmosphere lead to global warming[1]. $CO_2$ is one of the main anthropogenic greenhouse gasses. Accumulation of the atmospheric $CO_2$ result from industry-related consumption such a combustion of fossil-based fuels which are comprising 87% of the total energy consumption to provide World energy demand[2]. One of the sources of $CO_2$ emission is flue gas which is released power plants. It comprises of more than 7% of total $CO_2$ emission. Burning of the fossil-based fuels such as coal, oil and natural gas results in the flue gas which emits into the atmosphere[1]. It usually consists of the vast percentage of nitrogen, $CO_2$, water vapour and a small number of air pollutants (in ppm level) such as carbon monoxide (CO), nitrogen oxides ($NO_x$) and sulphur oxides ($SO_x$)[3].

In order to diminish $CO_2$ levels, two major ways can be applied: the chemical reaction-based and the biological mitigation approaches. The first way has high costly applications, excessively amount of energy consumption, and potential $CO_2$ leakage over time[3]. The biological ways are applied through photosynthesis by based on usage of light, an organic substrate such a glucose and an inorganic substrate such a $CO_2$ (phenomenon is called $CO_2$ fixation). Besides, photosynthesis is not only diminishing $CO_2$ level in the atmosphere but also promotes biomass biodiesel production[3].

Microalgae are microorganisms capable of converting sunlight into chemical energy through photosynthesis like plant do. They grow in wide range of aquatic habitats such as lakes, rivers, oceans, and even wastewater[4]. Microalgae can easily grow and proliferate with some fundamental requirements such as light, temperature, sugar, $CO_2$, nitrogen. Microalgae were able to grow photoautotrophically, heterotrophically, and mixotrophically. In many studies, photo-autotrophic growth is used in microalgae cultivation for $CO_2$ reduction because of low cost and ability to commercialize. Besides, microalgae have much higher growth rate, do not require any fertile land, require less land area than plants. Moreover, microalgae cells accumulate high amount of lipid content with utilization of direct flue gases as inorganic carbon source in form of $CO_2$[4,5].

Microalgae gain tremendous attention for reducing $CO_2$ in the last years[6]. Especially, species from genus *Chlorella* is most understood species in current research and has been considered promising microorganisms for $CO_2$ fixation. Numerous studies represent that flue gas could be used as a carbon source for $CO_2$ fixation and production of microalgae with *Chlorella*, without any adverse effects[3,6]. In the García-Cubero *et al.* study, *Chlorella vulgaris* has been grown direct flue gas feed. Besides, the amount of lipid and protein are increased with direct flue gas feed[3]. In the study of Borkenstein *et al.*, it was observed that *Chlorella emersonii* cultivated with flue gas feed grew without any side effects. It even positively affected its growth rate[7]. Microalgae cultivation with flue gas feed is a promising strategy for both reduction of $CO_2$ and production of the valuable biomass in economical way. As far as we aware, effect of $CO_2$ and flue gas on cell growth of microalgae *Chlorella protothecoides* has not been studied.

*Chlorella protothecoides* has a valuable lipid source that chemical composition is quite similar to vegetable oil[8]. The aim of this study was investigation of $CO_2$ fixation by photoautotrophic growth of microalgae in different concentrations of $CO_2$ and simulated flue gas for reducing $CO_2$ in the atmosphere and produce biomass. The growth kinetics, biomass productivity, lipid contents and lipid amount of *C. protothecoides* were analysed.

**Experimental**

*Materials:*

*Chlorella protothecoides* culture was purchased from Ege Biotechnology Inc. (İzmir, Turkey). All chemicals purchased from Sigma-Aldrich and Merck-Germany, were used without further any purification.

*Culture conditions of Chlorella protothecoides:*

In the present study, medium was prepared similar to our previous study[8]. Briefly, $NaNO_3$ (0.25 g), $K_2HPO_4$ (0.075 g), $KH_2PO_4$ (0.175 g), $MgSO_4.7H_2O$ (0.075 g) $CaCl_2.2H_2O$ (0.025 g), NaCl (0.025 g), and trace metal mixture was used as Bold Basal Medium (BBM) in order to preserve and cultivate *C. protothecoides* cells in 1 L. Trace metal mixture was prepared by using $H_3BO_3$ (11.42 g), $MnCl_2.4H_2O$ (1.44 g), $ZnSO_4.7H_2O$ (0.25 g), $CuSO_4.5H_2O$ (1.57 g), $Co(NO_3)_2.6H_2O$ (0.49 g), $MoO_3$ (0.71 g), EDTA (50.00 g), KOH (31.00 g), $FeSO_4.7H_2O$ (4.98 g) and $H_2SO_4$ (1 mL). Stock solutions of all chemical components of BBM were prepared and sterilized in autoclave (Systec 2540 EL,) at 121°C for 20 min. The initial pH values of the growth medium were adjusted to between 7.5–8. Microalgae cell cultivation was carried out at 28°C under 1000 lux fluorescence bulb (60 W) illumination with a 12:12 h light:dark regime. The precultures of *C. protothecoides* were grown in 250 mL erlenmeyer flasks containing BBM. To scale-up microalgae cells from preculture solution, they were transferred into 2 L culture bottles with an 20% (v/v) inoculum ratio. Cultures were continuously aerated with a compressor. Microalgal cells were cultivated in laboratory environment with air as a control medium, 10% and 15% $CO_2$ and simulated flue gas. Air flow rate was adjusted as 1 L/min with a flowmeter. And pH of the cultures was adjustedby pH controller. To see effect of $CO_2$, culture medium feed with air, 10% and 15% (v/v) $CO_2$ concentration. Cell growth was maintained for 12 days. To see effect of flue gas, *C. protothecoides* was incubated with simulated flue gas and 15% $CO_2$. The composition of the simulated flue gas was 75% $N_2$, 15% $CO_2$, 10% $O_2$, 400 ppm $SO_2$, and 125 ppm $NO_2$. Flow rate of flue gas was adjusted as 600 ml/min. Cell cultivation was maintained for 6 days.

*Harvesting of C. protothecoides biomass:*

Cell growth was monitored, and cultivation was maintained until cultures reach the stationary phase. After reaching stationary phase cells were harvested by 20 min centrifugation at 4800 rpm and cell pellet was washed three times with distilled water. Biomass was freeze-dried using freeze dryer. Dried biomass was kept at –20°C until further analysis.

*Determination of growth kinetics:*

Microalgal cell growth was monitored by daily changes in optical density at wavelength of 500 nm ($OD_{500}$) using UV/Vis spectrophotometer (Scinco S-3100, Seoul, Korea) and cell count was performed on the microalgal suspensions using an Olympos CX40 microscope. Microalgal dry cell weight (g/L) was measured three-day intervals. In order to calculate dry cell weight, 50 mL algae samples were taken and concentrated by filtering through Whatman GF 6 glass fiber filter paper (pore size: 1–3 µm) under vacuum. Cells were washed with pH:4 HCl solution, subsequently dried in pre-weighed petri plates in an oven at 105°C for 2 h. Dry cell weight (%) was determined gravimetrically. The following methods are described by Xia *et al.*[9]. The specific growth rate ($\mu$) is calculated by eq. (1) during the exponential phase. The biomass productivity (g/L/day) is determined using eq. (2).

$$\mu = \left(\ln\frac{X_2}{X_1}\right) \times \frac{1}{t} \quad (1)$$

$$P = \left(\frac{W_e - W_b}{V \times d}\right) \times 100\% \quad (2)$$

where $X_1$ and $X_2$ are the biomass at time 1 ($t_1$) and time 2 ($t_2$), respectively; $W_b$ and $W_e$ are the harvested cell dry mass at the beginning and end of cultivation, respectively; $P$ is productivities of microalgal biomass; $V$ and $d$ are culture volume ($L$), and time of culture ($d$), respectively.

*Lipid analysis:*

Lipid extraction was performed at the end of exponential phase using the following procedure: 3 ml hexane as a solvent was added to each 0.05 g dried algal biomass since hexane has ability to solve lipid molecules. Then microalgae cells were homogenized using a homogenizer and incubated in a water bath for 24 h at 25°C for lipid extraction. Hexane phase including lipid molecules was separated from cell debris by centrifugation for 20 min at 4000 rpm. Hexane was evaporated at 60°C and lipid content was determined gravimetrically.

**Results and discussion**

In this study, the effects of different $CO_2$ concentrations and simulated flue gas feed on the growth *C. protothecoides* were investigated at the constant temperature, pH and light intensity. Growth kinetics, biomass productivity, and lipid content of *C. protothecoides* were determined.

*Effect of the different concentration $CO_2$:*

*C. protothecoides* was cultivated with air and two different concentration $CO_2$ feed. Fig. 1 shows the comparison of *C. protothecoides* feed with air (control), 10% and 15% $CO_2$ concentration during a 12-day trial period. The maximum cell density was reached after 12 days. The results have been quite evidence for the effect of $CO_2$ on the growth of *C. protothecoides*. The results indicate that *C. protothecoides* culture feed with 15% $CO_2$ at the end of the 12th day reached the highest density with an $OD_{500}$ of 3.42 value. 10% $CO_2$ concentration $OD_{500}$ value is also higher than air. These data suggest that using a high level of $CO_2$ increases obviously *C. protothecoides* biomass concentration. It has been observed that the highest dry cell weight reached with 10% $CO_2$ level (Fig. 2). The highest biomass productivity and the specific growth rate were found 0.125 g/L/day and 1,10 $\mu$, at 10% $CO_2$ concentration, respectively (Table 1). In the study of Hanagata *et al.*, *Scenedesmus sp.* and *Chlorella sp.*, for the two organisms, the biomass productivity was observed 0.15 g/L/day and 0.18 g/L/day at 10% and 45% $CO_2$ concentrations, respectively[10]. In our study, it was observed that *C. protothecoides* had high tolerance to $CO_2$ and positively affected dry cell weight and biomass productivity. It can be successful in large-scale production due to its high biomass productivity.

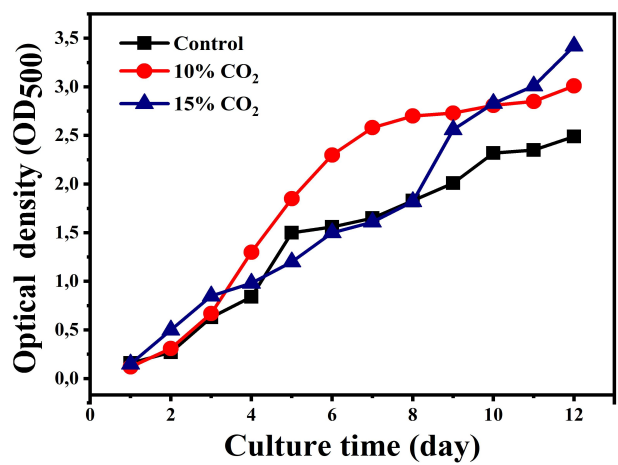

Fig. 1. Optical density values of *C. protothecoides* with control and different concentrations $CO_2$ feed.

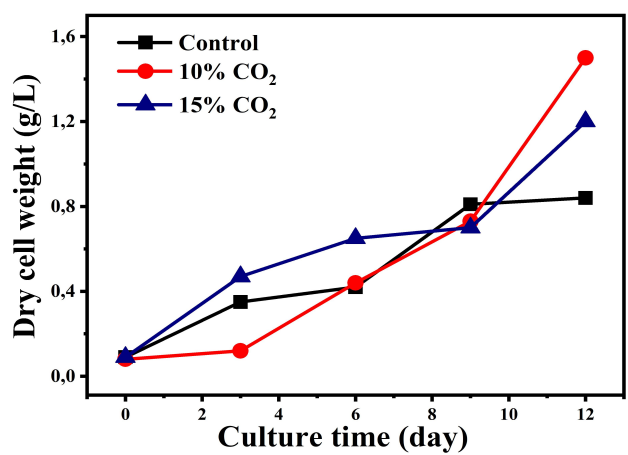

Fig. 2. Dry cell weight (g/L) of *C. protothecoides* with control and different concentrations $CO_2$ feed.

Table 1. Growth kinetics and lipid amount of *C. protothecoides* in different feed

| Sample | Dry cell weight (g/L) | Biomass productivity (g/L/day) | Specific growth rate ($\mu$) | Lipid weight (g lipid/g biomass) | Lipid content (%) |
|---|---|---|---|---|---|
| Air (Control) | 0,84 | 0,07 | 0,77 | 0,31 | 38 |
| 10% $CO_2$ | 1,5 | 0,125 | 1,10 | 0,63 | 42 |
| 15% $CO_2$ | 1,2 | 0,1 | 0,84 | 0,54 | 45 |

Microalgae resist to adapt to stresses and altering conditions in their environment. Lipids can act as signal components, structural components of the plasma membranes of the cell[11]. Vargas *et al.* reported that stress conditions increase the amount of lipid. Increasing the concentration of $CO_2$ in some species may affect the accumulation of lipid as it may cause lower pH[11]. Sun *et al.* reported that increasing concentration of $CO_2$ enhance lipid accumulation by using *Chlorella sorokiniana*[14]. In our study, 15% $CO_2$ resulted higher lipid content than 10% $CO_2$ (Fig. 3a). It has been seen that increased lipid content increases with increasing $CO_2$, which is in parallel with the study of the Xia *et al.* on the *C. Sorokiniana* species[9]. The highest level of lipid amount (0,63 g/L) was reached at 10% $CO_2$ (Fig. 3b). Even if it seems conflict, this result comes from high dry cell weight of 10% $CO_2$.

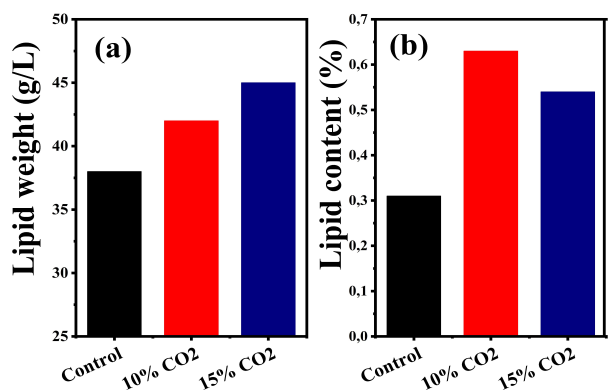

Fig. 3. Lipid weight (a) and content (b) of *C. protothecoides* at different $CO_2$ levels.

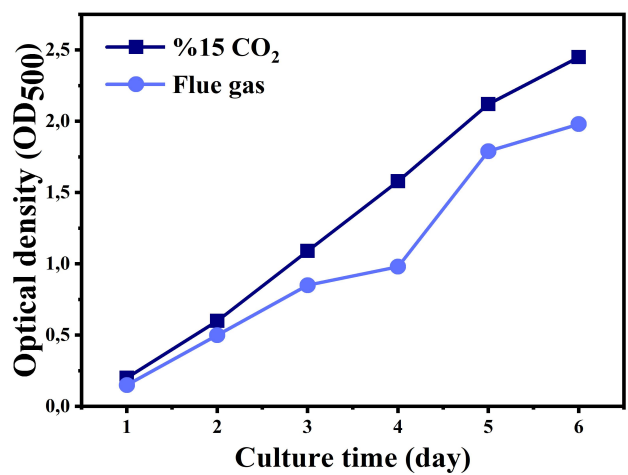

Fig. 4. Optical density values of *C. protothecoides* fed by flue gas and 15% $CO_2$ in 6-day interval.

*Effect of the simulated flue gas:*

Flue gas injection to the culture mediums cause the rise of temperature, and also flue gas includes of $SO_x$ and $NO_x$ compounds which are toxic for microalgae cells[13]. Microalgae culture that was feed with flue gas reached the stationary phase after 5th day whereas, feed with 15% $CO_2$ reached the stationary phase after 6 days and its cell count was found similar (Fig. 5). Flue gas and 15% $CO_2$ feed show very similar dry cell weight values (Table 2). These results implicate that growth of *C. protothecoides* was not inhibited by flue gas and could tolerate 125 ppm $NO_2$ and 400 ppm $SO_2$.

Specific growth rate and biomass productivty of culture which is feed with 15% $CO_2$ was slightly higher than those observed with flue gas feed (Table 2). It can be said that flue gas did not inhibited growth rate. That may be due to the drop of medium pH because of $NO_x$ and $SO_2$ content of flue gas. When $CO_2$ is dissolved in water aqueous $CO_2$, carbonic acid ($H_2CO_3$), bicarbonate ($HCO_3^-$), carbonate ($CO_3^{2-}$) ions, which affect the pH of medium, are formed. With the addition of flue gas in the culture media, depending on the pH, tem-

Table 2. Growth kinetics and lipid amount of *C. protothecoides* in 15% $CO_2$ and simulated flue gas feed

| Sample | Dry cell weight (g/L) | Biomass productivity (g/L/day) | Specific growth rate ($\mu$) | Lipid content (%) | Lipid weight (g/g) |
|---|---|---|---|---|---|
| 15% $CO_2$ | 0,65 | 0,108 | 0,71 | 45 | 0,54 |
| Flue gas | 0,62 | 0,103 | 0,65 | 44 | 0,27 |

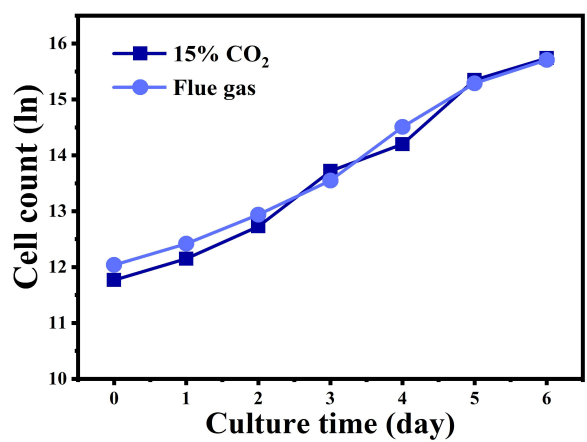

**Fig. 5.** Specific growth rate of *C. protothecoides* fed by 15% $CO_2$ and flue gas.

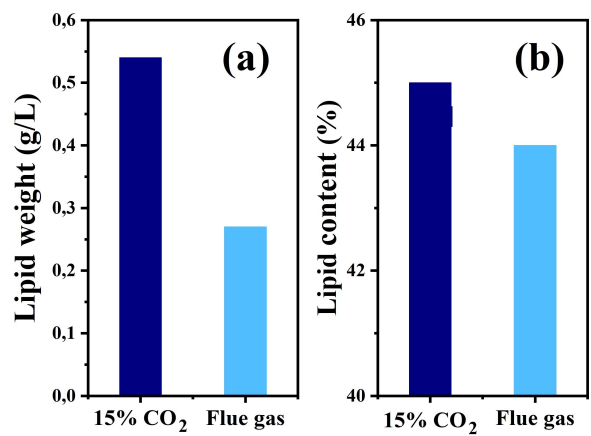

**Fig. 6.** Lipid weight (a) and content (b) of *C. protothecoides* fed by 15% $CO_2$ and flue gas feed.

perature and pressure of solution the following reaction takes place:

$$CO_{2(aq)} + H_2O \leftrightarrow H_2CO_3 \leftrightarrow H^+ + HCO_3^- H \leftrightarrow 2H^+ + CO_3^{2-} \quad (3)$$

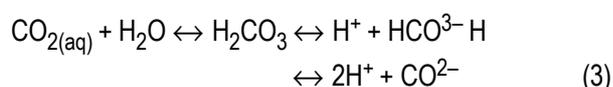

On the other hand, during cell growth microalgae cells utilize $CO_2$ as carbon source. Thereby, equilibrium shifts towards reactants and pH of the medium is increased.

In the present study, effect of flue gas on lipid weight and lipid content of *C. protothecoides* were also investigated and represented in Fig. 6. It was seen that lipid amount of microalgae cells cultured with 15% $CO_2$ was about two-fold of lipid amount produced by the culture feed with flue gas. On the other hand, lipid content (%) was almost same for 15% $CO_2$ (45%) and simulated flue gas (44%) feed. The results suggest that flue gas can be utilized as an alternative carbon source for biomass and lipid production of *C. protothecoides*. Thus, flue gas emission to the atmosphere may be reduced along with microalgae biomass production which could huge potential energy source such a biodiesel.

## Conclusions

*C. protothecoides* was cultured in the air as a control, 10%, 15% $CO_2$ and flue gas (15% $CO_2$). Growth kinetics, lipid content and weight were investigated. Increasing of $CO_2$ level have an incontrovertible positive effect on dry cell weight lipid content, and lipid weight of *C. protothecoides*. Results supported that *C. protothecoides* can tolerate the flue gas as well as a similar growth rate with $CO_2$. This means that direct flue gas can be used as a carbon source. All these results have shown that *C. protothecoides* can be a potential source in reducing $CO_2$ emissions. Lipid recovered from *C. protothecoides* with flue gas feed can a promising candidate for further biodiesel applications because of low costs and positive environmental effect.

## Acknowledgements

The authors are grateful to the Yildiz Technical University Projects Office for the financial support through an extensive research project (Project No. 2015-07-04-KAP-03).

**References**

1. (a) J. Davidson, P. Freund and A. Smith, "Putting Carbon Back in the Ground", IEA, 2001, pp. 1-30; (b) D. Aaron and C. Tsouris, *Sep. Sci. Technol.*, 2005, **40**, 321.

2. (a) B. Petrolum, "BP Statistical Review of World Energy", 2012; (b) K. Kumar, C. N. Dasgupta, B. Nayak, P. Lindblad and D. Das, *Bioresour. Technol.,* 2011, **102**, 4945.

3. (a) P. Kandimalla, S. Desi and H. Vurimindi, *Environ. Sci. Pollut. Res.,* 2016, **23**, 9345; (b) R. García-Cubero, J. Moreno-Fernández and M. García-González, *Waste Biomass. Valor.,* 2018, **9**, 2015; (c) B. Wang, Y. Li, N. Wu and C. Lan, *Appl. Microbiol. Biotechnol.*, 2008, **79**, 707.

4. (a) M. Khan, J. Shin and J. Kim, *Microb. Cell. Fact.,* 2018, **17(36)**; (b) Y. Chisti, *Biotechnol. Adv.,* 2007, **25**, 294.

5. A. M. Kunjapur and R. B. Eldridge, *Ind. Eng. Chem. Res.,* 2010, **49**, 3516.

6. (a) J. Doucha, F. Straka and K. Lívanský, *Appl. Phycol.,* 2005, **17**, 403; (b) K. Kadam, *Energy Convers. Manag.,* 1997, **38**, 505; (c) R. Ramanan, K. Kannan, A. Deshkar, R. Yadav and T. Chakrabarti, *Bioresour. Technol.,* 2010, **101(8)**, 2616.

7. C. Borkenstein, J. Knoblechner, H. Frühwirth and M. Schagerl, *J. Appl. Phycol.*, 2011, **23**, 131.

8. S. Yucel, P. Terzioglu, M. E. Bogoclu and M. Celikkol, *Sigma J. Eng. & Nat. Sci.,* 2016, **34**, 183.

9. J. Xia, S. Gong, X. Jin, M. Wan and Z. Nie, *J. Cent. South. Univ.,* 2013, **20**, 730.

10. N. Hanagata, T. Takeuchi, Y. Fukuju, D. J. Barnes and I. Karube, *Phytochemistry,* 1992, **31**, 3345.

11. L. Barsanti, and P. Gualtieri, "Algae: Anatomy, Biochemistry, and Biotechnology", CRC Press, 2005, Chap. 7, 267-309.

12. Z. Sun, X. Dou, J. Wu, B. He, Y. Wang and Y. F. Chen, *World J. Microbiol. Biotechnol.,* 2016, **32**, 9.

13. H. Matsumoto, A. Hamasaki, N. Sioji and Y. Ikuta, *J. Chem. Eng. Jpn*., 1997, **30**, 620.